# Surface and bulk effects of K in highly efficient $Cu_{1-x}K_xInSe_2$ solar cells

Christopher P. Muzzillo, Jian V. Li, Lorelle M. Mansfield, Kannan Ramanathan, and Timothy J. Anderson*

Dr. C.P. Muzzillo, Prof. T.J. Anderson

Department of Chemical Engineering, University of Florida, Gainesville, FL, 32611, USA

E-mail: tim@ufl.edu

Dr. L.M. Mansfield, Dr. K. Ramanathan

National Renewable Energy Laboratory, 15013 Denver West Pkwy, Golden, CO, 80401, USA

Prof. J.V. Li

Department of Physics, Texas State University, San Marcos, TX, 78666, USA

Abstract:

To advance knowledge of K bonding in $Cu(In,Ga)(Se,S)_2$ (CIGS) photovoltaic (PV) absorbers, recent Cu-K-In-Se phase growth studies have been extended to PV performance. First, the effect of distributing K throughout bulk $Cu_{1-x}K_xInSe_2$ absorbers at low K/(K+Cu) compositions ($0 \leq x \leq 0.30$) was studied. Efficiency, open-circuit voltage ($V_{OC}$), and fill factor (FF) were greatly enhanced for x ~ 0.07, resulting in an officially-measured 15.0%-efficient solar cell, matching to the world record $CuInSe_2$ efficiency. The improvements were a result of reduced interface and bulk recombination, relative to $CuInSe_2$ (x ~ 0). However, higher x compositions had reduced efficiency, short-circuit current density ($J_{SC}$), and FF due to greatly increased interface recombination, relative to the x ~ 0 baseline. Next, the effect of confining K at the absorber/buffer interface at high K/(K+Cu) compositions ($0.30 \leq x \leq 0.92$) was researched. Previous work showed that these surface layer growth conditions produced $CuInSe_2$ with a large phase fraction of $KInSe_2$. After optimization (75 nm surface layer with x ~ 0.41), these $KInSe_2$ surface samples exhibited increased efficiency (officially 14.9%), $V_{OC}$, and FF as a result of decreased interface recombination. The $KInSe_2$ surfaces had features similar to previous reports for KF post-deposition treatments (PDTs) used in world record CIGS solar cells—taken as indirect evidence that $KInSe_2$ can form during these PDTs. Both the bulk and surface growth processes greatly reduced interface recombination. However, the $KInSe_2$ surface had higher K levels near the surface, greater lifetimes, and increased inversion near the buffer interface, relative to the champion bulk CKIS absorber. These characteristics demonstrate that K may benefit PV performance by different mechanisms at the surface and in the absorber bulk.

Graphical abstract:

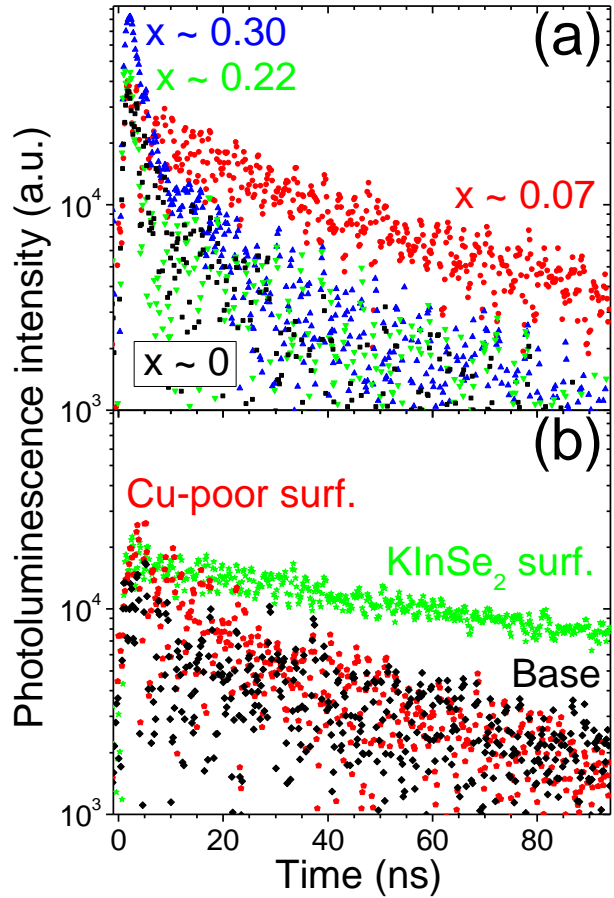

1. Introduction

Recent reports have detailed power conversion efficiency enhancements when potassium fluoride and selenium were co-evaporated onto Cu(In,Ga)(Se,S)$_2$ (CIGS) absorbers at around 350°C (KF post-deposition treatment (PDT)).[1-20] RbF has also been used.[3, 17, 18] In particular, 6 of the last 8 world record CIGS efficiencies have employed a KF (or RbF) PDT,[1, 3, 4, 12, 17, 21] ultimately advancing the record efficiency from 20.3 to 22.6% in just ~3.5 yr. KF PDT successes in the laboratory have now been extended to commercially-relevant chalcogenized CIGS absorbers,[12] full size (0.75 m$^2$) modules,[13] and Cd-free Zn(O,S) buffers.[2, 12, 22] Although the mechanisms responsible for these efficiency improvements are not clear, the KF PDT has been associated with multiple phenomena: *increased* hole concentration (e.g., by consuming In$_{Cu}$ compensating donors to produce K$_{Cu}$ neutral defects[23]),[5, 7, 8, 11, 14, 15, 19, 24-27] *decreased* hole concentration (by consuming Na$_{Cu}$ to produce In$_{Cu}$ compensating donors,[1] or by forming (K-K)$_{Cu}$ dumbbell interstitial donors[28]),[1, 8, 10, 16] Na depletion or formation of soluble Na chemical(s),[1, 5, 7, 8, 10, 13, 14, 25, 26, 29, 30] Ga depletion at the surface,[1, 8, 10, 13-15, 29, 31, 32] Cu depletion at the surface[7, 13, 15, 18, 20] resulting in better near-surface inversion[1, 8, 10, 16, 33] or decreased valence band energy,[14, 16, 24, 31, 32, 34] grain boundary passivation,[5, 12, 20, 35] general defect passivation,[2, 12, 26] minority carrier lifetime enhancement,[9, 15, 19] decreased minority carrier diffusion length,[19] more Cd diffusion into the absorber (or a *less* abrupt interface),[1, 2, 8-10, 14, 24, 33] a *more* abrupt absorber/buffer interface,[16] morphology changes resulting in increased CdS nucleation sites,[2, 10, 36] general changes in CdS growth,[14, 31, 33, 37] formation of a passivating K-In-Ga-Se,[10, 38] K-In-Se,[14, 18, 31, 32, 37, 39] or In-Se[14, 18, 32, 37, 39] interfacial compound, formation of a current blocking interfacial compound,[25] formation of elemental Se at the surface,[14, 37, 39] formation of surface Cu$_{2-x}$Se and GaF$_3$,[14] increased formation of surface In-O and/or Ga-O after air exposure,[14, 20, 24, 31, 34] consumption of a surface 'ordered vacancy compound,'[14] decreased trap concentration,[19, 29] reduced microscopic fluctuations in surface resistivity and potential,[16] and reduced nanoscopic fluctuations in potential.[19] Substrate surfaces with higher K content also led to CIGS with increased carrier concentrations,[6, 23, 40] modified Cu-Ga-In interdiffusion,[6, 40] as well as a K-rich and Cu-poor surface.[41] Similarly, an RbF PDT was associated with general defect passivation,[3] and a CsF PDT caused Na and K depletion.[17] Another group co-evaporated KF, In, and Se onto CIGS, possibly forming KInSe$_2$ or K-doped amorphous In$_2$Se$_3$, and achieving results similar to a KF PDT.[37, 39] A PDT without KF (just Se) has also been shown to significantly alter CIGS absorbers,[42] calling into question what 'control' absorber is appropriate for KF PDT comparison. High absorber Na and K composition has also been linked to drastically accelerated degradation in photovoltaic (PV) performance,[43] underscoring the importance of understanding alkali metal bonding in CIGS.

While the mechanisms underlying KF PDT effects remain uncertain, it has been established that a relatively large amount of K is present at the p-n junction in the most efficient solar cells.[1] That finding has heightened interest in Cu-K-In-Se material with group I-poor (i.e. (K+Cu)/In ~ 0.85) and K-rich (x > 0.30) compositions.[30, 44, 45] Studies of Cu-K-In-Se phase growth at those compositions were recently initiated: Cu, KF, In, and Se were co-evaporated to form $Cu_{1-x}K_xInSe_2$ (CKIS) alloys with K/(K+Cu) composition, or x, varied from 0 to 1.[30, 44-46] Increasing x in CKIS was found to monotonically decrease the chalcopyrite lattice parameter, increase the band gap, and increase the apparent carrier concentration. Moderate K compositions (0 < x < 0.30) exhibited significantly longer minority carrier lifetimes,[44-46] relative to x ~ 0 and x ≥ 0.30. Superior PV performance was also observed for x ~ 0.07 for a Ga-alloyed film with Ga/(Ga+In) ~ 0.3.[26] The substrate surface's Na composition[30] and temperature[45] were found to determine the relative amount of CKIS alloy formation during growth, relative to $CuInSe_2$ + $KInSe_2$ mixed-phase formation. These advances in using processing conditions to determine K bonding in chalcopyrite films have been presently extended to PV performance. In this work, the effect of K on PV performance was examined for two cases: (1) where K is distributed throughout the bulk absorber at low K/(K+Cu) compositions, and (2) where K is confined to the absorber/buffer interface at high K/(K+Cu) compositions. These configurations were chosen to help distinguish surface and bulk effects of K in chalcopyrite films, where both may have influenced successful KF PDT results.[1-20]

2. Results and discussion

2.1. Low K/(K+Cu) throughout bulk absorber

Bulk CKIS absorbers with K/(K+Cu), or x ~ 0, 0.07, 0.22, and 0.30 were grown on soda-lime glass (SLG)/Mo substrates at 500°C, as seen in Figure S1. X-ray diffraction (XRD) yielded smaller lattice parameters at increased x, taken as evidence of CKIS alloy formation and in line with a previous report.[44] The formation of CKIS alloys was confirmed by vacuum annealing films to 600°C (10 min; Se provided). After annealing, films with x ≥ 0.07 had a small but reproducible shift to larger lattice parameters that was not observed in the baseline x ~ 0 films, and indicated CKIS alloy decomposition (e.g. Figure S2).[45] The changes in PV performance with x are therefore attributed to CKIS alloying (Figure 1), but minor $KInSe_2$ formation could also play a role.[30, 45] The x ~ 0.07 film had increased open-circuit voltage ($V_{OC}$) and fill factor (FF), leading to higher efficiency (relative to x ~ 0). Further increasing x to 0.22 and 0.30 caused a reduction in short-circuit current density ($J_{SC}$) and FF, while $V_{OC}$ remained constant, resulting in lower efficiency. The films with larger x sometimes delaminated from the Mo film on humid air exposure or during chemical bath deposition (CBD). This was avoided with films < 1 μm in thickness,[44] but solar cells were not fabricated with such thin absorbers. External quantum efficiency (QE) in Figure S3 showed very similar collection for the champion x ~ 0 and 0.07 absorbers. Absorbers with x ~ 0.22 and 0.30 had decreased collection—particularly at long wavelengths, which could be a result of lower diffusion lengths (e.g. through lower lifetimes in Table S1; Figure 2). For the poor-performing x ~ 0.22 and 0.30 absorbers, the long wavelength QE cutoffs did not correlate with band gaps estimated from XRD. The difference in apparent CdS thickness observable in Figure S3 was incidental, and no trend in apparent CdS thickness or CdS collection was found for CKIS absorbers. The champion performance is summarized in Table 1. An anti-reflective film was applied to the x ~ 0.07 champion, and it was officially measured at 15.0% efficient (Figure S4), matching the world record efficiency for a $CuInSe_2$-based solar cell.[47] The present study's 15.0%-efficient absorber was grown at lower temperature by ~75°C, did not use the 'gold standard' three stage process,[47] and no attempt was made at optimizing the device stack around it, so even higher efficiencies are likely.

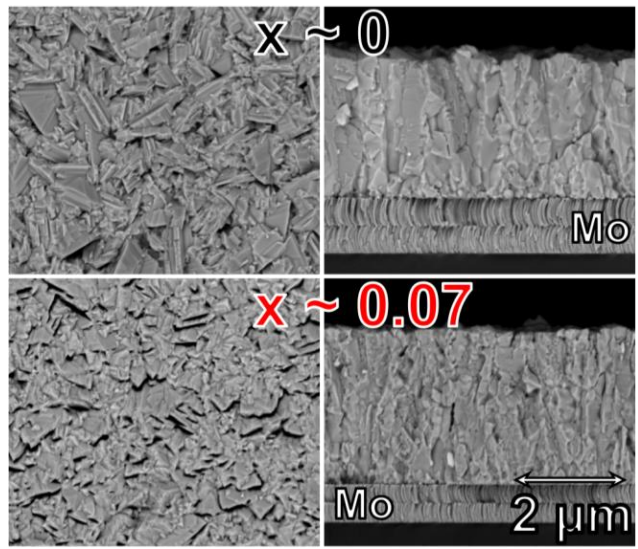

Figure S1. Plan view (left) and cross-sectional (right) SEM micrographs of SLG/Mo/CKIS films with K/(K+Cu), or x ~ 0 (top) and 0.07 (bottom) grown at 500°C.

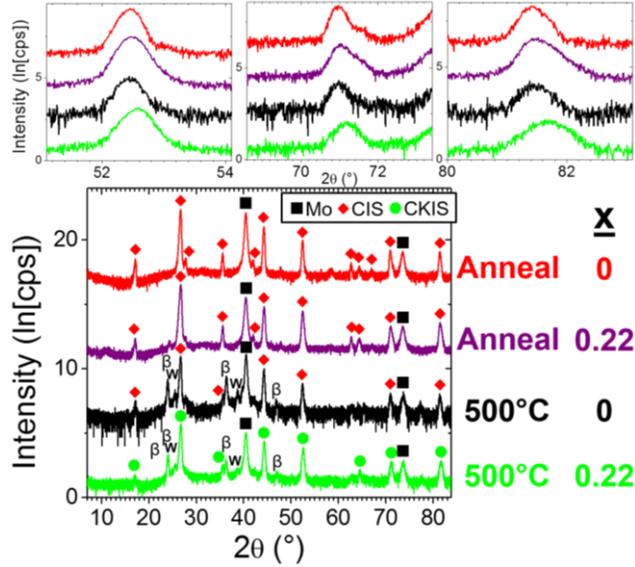

Figure S2. XRD scans from Cu-K-In-Se films with K/(K+Cu), or x ~ 0 (3rd; black) and x ~ 0.22 (bottom; green) grown on SLG/Mo at 500°C, with the x ~ 0 (top; red) and x ~ 0.22 (2nd; purple) then annealed at 600°C for 10 min. Mo, CuInSe$_2$, and CKIS peaks are labeled with black squares, red diamonds, and green circles, respectively. 'β' and 'W' peaks are from Cu K$_\beta$ and W impurity in the Cu radiation source.

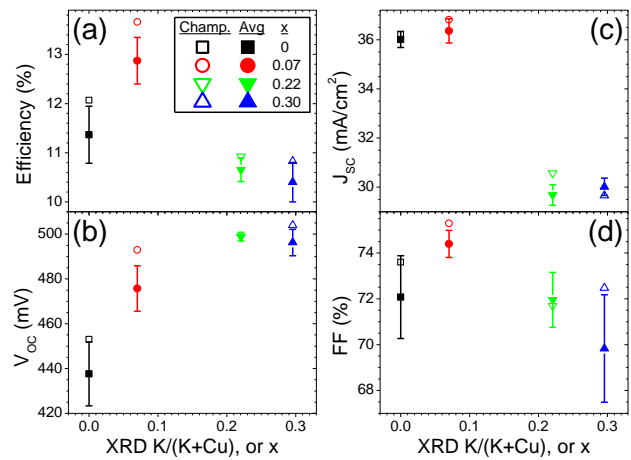

Figure 1. Champion (empty symbols) and average (filled symbols) efficiency (a), open-circuit voltage ($V_{OC}$) (b), short-circuit current density ($J_{SC}$) (c), and fill factor (FF) (d) versus K/(K+Cu), or x composition by XRD. Standard deviations are shown as error bars.

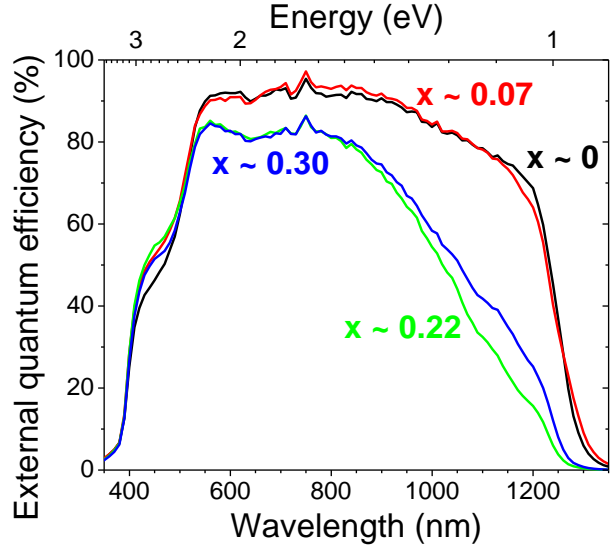

Figure S3. External QE for CKIS absorbers with x ~ 0 (black), 0.07 (red), 0.22 (green), and 0.30 (blue) grown at 500°C.

Table S1. Maximum lifetime ($\tau_{single,max}$) and carrier concentration ($p_{TRPL,max}$) extracted from TRPL on bare absorbers, and defect density ($N_{CV}$), depletion width ($W_d$), and built-in voltage ($V_{bi}$) extracted from CV on best devices made with those absorbers. The asterisk (*) denotes values with low confidence.

| Growth temp. (°C) | 500 | | | | 600 | | |
|---|---|---|---|---|---|---|---|
| Sample | Bulk x ~ 0 | Bulk x ~ 0.07 | Bulk x ~ 0.22 | Bulk x ~ 0.30 | Base | Cu-poor surf. | KInSe$_2$ surf. |
| $\tau_{single,max}$ (ns) | 5 | 23 | 3 | 4 | 8 | 14 | 64 |
| $p_{TRPL,max} \cdot 10^{15}$ (cm$^{-3}$) | 4 | 4 | 7 | 11 | 3 | 3 | 6 |
| $N_{CV} \cdot 10^{16}$ (cm$^{-3}$) | 0.9 | 4.6 | 7.5 | 7.2 | 2.8 | 3.7 | 5.1 |
| $W_d$ (nm) | 350 | 220 | 190 | 190 | 220 | 230 | 180 |
| $V_{bi}$ (mV) | 662 | 990 | 1275* | 1125* | 749 | 873 | 977 |

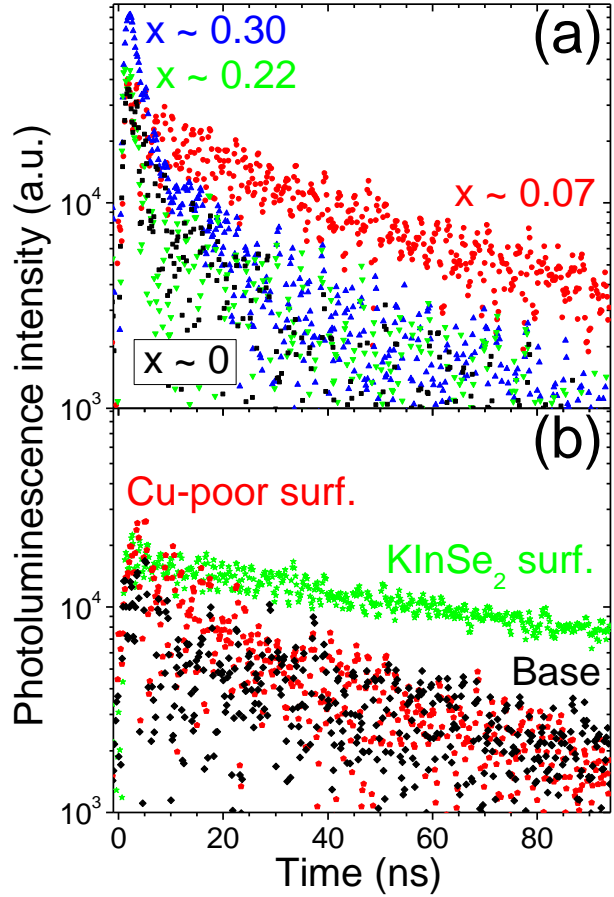

Figure 2. TRPL decays for (a) bulk CKIS (grown at 500°C) with x ~ 0 (black), x ~ 0.07 (red), x ~ 0.22 (green), and x ~ 0.30 (blue) and (b) 600°C base (black), Cu-poor surface (red), and KInSe$_2$ surface (green) absorbers.

Table 1. Band gap ($E_g$), $V_{OC}$ extrapolated to 0 K ($V_{OC}(T=0\ K)$), AS activation energy ($E_{a,AS}$), power conversion efficiency, open-circuit voltage ($V_{OC}$), short-circuit current density ($J_{SC}$), fill factor (FF), and interface, depletion region, and bulk recombination rates ($R_{interf.}$, $R_{depl.}$, and $R_{bulk}$) and resulting photovoltage deficits ($D_{interf.}$, $D_{depl.}$, and $D_{bulk}$; defined in Equation (1)). The values in parenthesis are official measurements after anti-reflective coating was applied.

| Growth temp. (°C) | 500 | | | | 600 | | |
|---|---|---|---|---|---|---|---|
| Sample | Bulk x ~ 0 | Bulk x ~ 0.07 | Bulk x ~ 0.22 | Bulk x ~ 0.30 | Base | Cu-poor surf. | KInSe$_2$ surf. |
| $E_g$ (eV) | 1 | 1 | 1.04 | 1.10 | 1 | 1 | 1 |
| $V_{OC}(T=0\ K)$ (V) | 0.99 | 1.04 | 1.02 | 1.04 | 0.98 | 1 | 1.03 |
| $E_{a,AS}$ (meV) | 89 | 70 | 100 | 70 | 77 | 87 | 31 |
| Efficiency (%) | 12.1 | 13.7 (15.0) | 10.9 | 10.8 | 12.1 | 12.6 | 14.2 (14.9) |
| $V_{OC}$ (mV) | 453 | 493 (506) | 499 | 502 | 448 | 463 | 491 (490) |
| $J_{SC}$ (mA/cm$^2$) | 36.2 | 36.8 (39.2) | 30.6 | 30.2 | 38.8 | 37.1 | 38.5 (40.8) |
| FF (%) | 73.6 | 75.3 (75.5) | 71.7 | 71.5 | 69.4 | 73.4 | 75.1 (74.6) |
| $R_{interf.}$ (cm$^{-2}$s$^{-1}$) | $2.5 \cdot 10^{16}$ | 0 | $2.9 \cdot 10^{17}$ | $3.5 \cdot 10^{17}$ | $6.4 \cdot 10^{16}$ | $5.1 \cdot 10^{16}$ | 0 |
| $D_{interf.}$ (mV) | -19 | 0 | -305 | -352 | -71 | -45 | 0 |
| $R_{depl.}$ (cm$^{-2}$s$^{-1}$) | $1.4 \cdot 10^{14}$ | $3.2 \cdot 10^{15}$ | $1.0 \cdot 10^{14}$ | $3.7 \cdot 10^{14}$ | $2.7 \cdot 10^{15}$ | $1.7 \cdot 10^{14}$ | $2.6 \cdot 10^{15}$ |
| $D_{depl.}$ (mV) | 0 | -2 | 0 | 0 | -3 | 0 | -3 |
| $R_{bulk}$ (cm$^{-2}$s$^{-1}$) | $3.9 \cdot 10^{17}$ | $3.8 \cdot 10^{17}$ | $4.2 \cdot 10^{14}$ | $6.0 \cdot 10^{15}$ | $2.2 \cdot 10^{17}$ | $2.9 \cdot 10^{17}$ | $2.5 \cdot 10^{17}$ |
| $D_{bulk}$ (mV) | -295 | -272 | 0 | -6 | -245 | -254 | -273 |

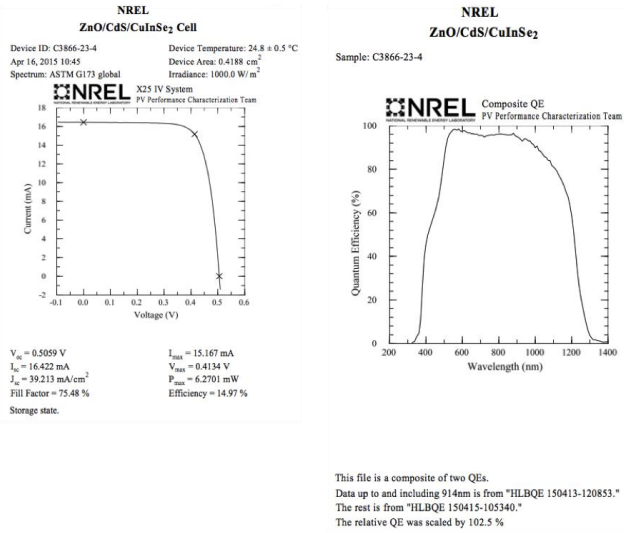

Figure S4. Officially measured current-voltage and external QE data for the x ~ 0.07 champion device.

Capacitance-voltage (CV), temperature- and illumination-dependent current density-voltage JV(T,G), and admittance spectroscopy (AS) were performed on the best devices. CV showed that x ~ 0, 0.07, 0.22, and 0.30 had increasing concentrations of ionized defects with increasing x (Figure S5; Table S1). These data agreed well with peak PL response (Table S1; Figure 2), as well as previous PL and current-voltage measurements.[44] Therefore, the ionized defects likely corresponded to shallow acceptors, with concentrations approximately equal to hole concentrations. The depletion width at zero bias and built-in voltage of each device were extracted from CV data, and are shown in Table S1.

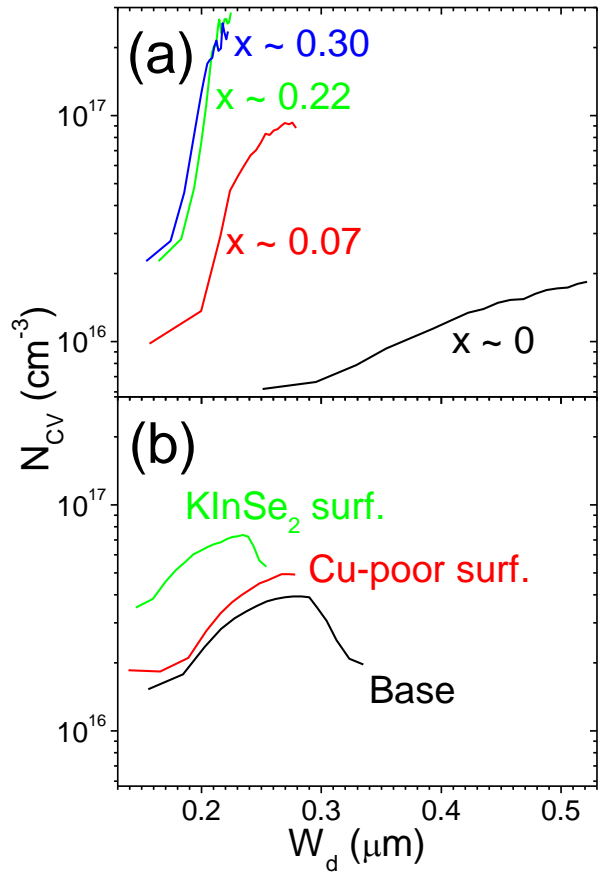

Figure S5. CV defect density versus depletion width ($W_d$) for (a) bulk CKIS (grown at 500°C) with $x \sim 0$ (black), $x \sim 0.07$ (red), $x \sim 0.22$ (green), and $x \sim 0.30$ (blue) and (b) 600°C base (black), Cu-poor surface (red), and KInSe$_2$ surface (green) absorbers.

At low temperature, all measured devices had JV rollover when biased above $V_{OC}$ at ~1 sun illumination (Figure S6), which is indicative of series resistance, or a current blocking barrier near the front[48] or back[8, 49, 50] interface of the absorber. All K-containing samples had less severe rollover. This behavior was also reported for a KF PDT, and was taken as evidence that K reduces a back contact barrier.[8] This could also relate to the apparent resistivity decrease in CKIS with increasing x composition.[44, 45]

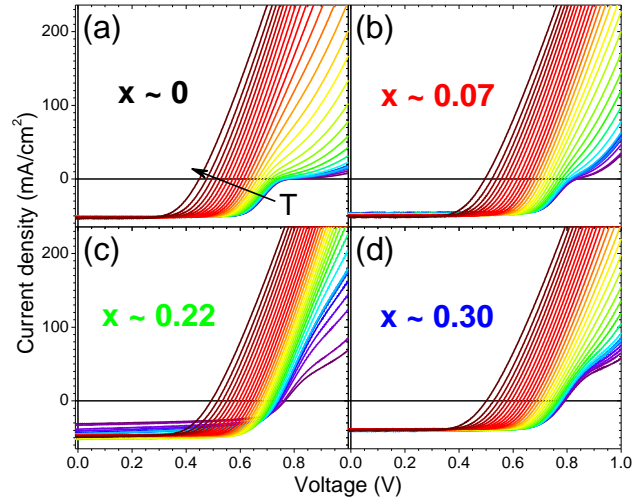

Figure S6. Temperature dependent JV at ~1 sun illumination for CKIS absorbers with (a) x ~ 0 (82 – 301 K), (b) x ~ 0.07 (89 – 301 K), (c) x ~ 0.22 (81 – 300 K), and (d) x ~ 0.30 (68 – 302 K). Higher temperature curves are more red; lower temperature curves are more violet.

CKIS films with x ~ 0.22 and 0.30 exhibited rollover in the $V_{OC}(T)$ data at relatively high temperatures, relative x ~ 0 and 0.07 (Figure 3). The extracted activation energies for recombination with no thermally generated carriers ($V_{OC}$(T=0 K) in Table 1) are a relative measure of interface and bulk recombination.[51] The most efficient device (x ~ 0.07) rolled over more strongly in $V_{OC}(G)$ at low illumination (Figure S7), which is related to relative interface, depletion region, and bulk region recombination rates.[52]

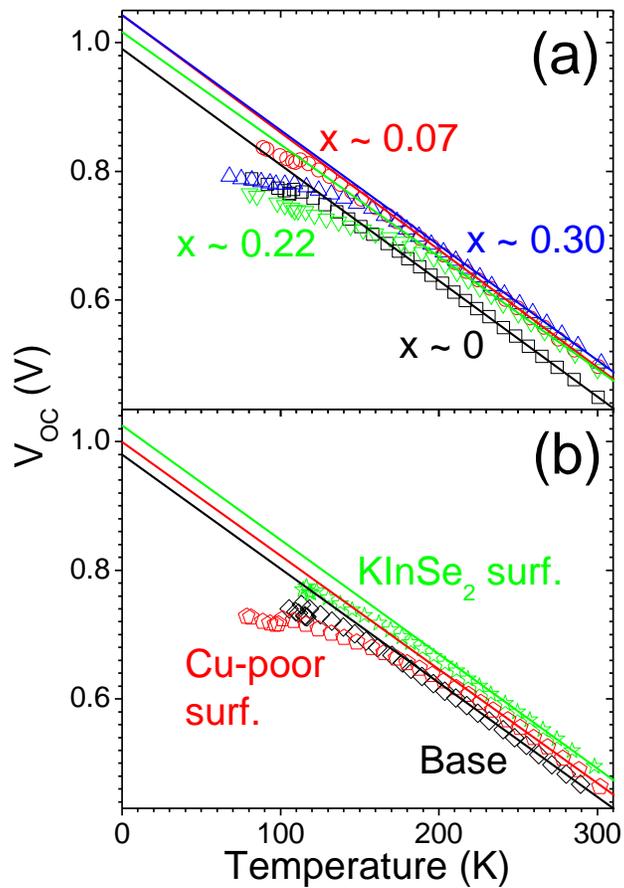

Figure 3. Open-circuit voltage versus temperature for (a) bulk CKIS (grown at 500°C) with x ~ 0 (black), x ~ 0.07 (red), x ~ 0.22 (green), and x ~ 0.30 (blue) and (b) 600°C base (black), Cu-poor surface (red), and KInSe$_2$ surface (green) absorbers. Lines are high temperature data fits.

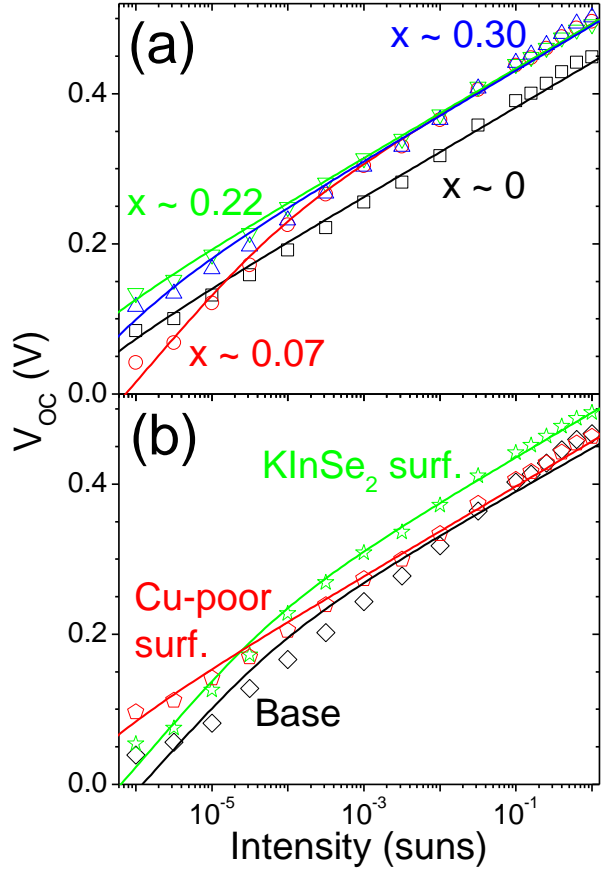

Figure S7. Open-circuit voltage versus illumination intensity for (a) bulk CKIS (grown at 500°C) with x ~ 0 (black squares), x ~ 0.07 (red circles), x ~ 0.22 (green down triangles), and x ~ 0.30 (blue up triangles) and (b) 600°C base (black diamonds), Cu-poor surface (red pentagons), and KInSe$_2$ surface (green stars) absorbers. Lines are two-parameter fits to the data.

AS revealed one signature in each sample (Figure S8). The activation energies of these signatures were 31 - 100 meV (Figure 4 and Table 1). Attempt-to-escape frequencies showed poor correlation with activation energies (Figure S9), but more data is needed to confirm this deviation from the Meyer-Neldel rule. Similar AS behavior was previously attributed to $E_c$-$E_F$ near the absorber/buffer interface,[53] bulk defects,[54] or a back contact barrier.[50] On the other hand, several phenomena can cause low temperature JV rollover,[8] but it is often assumed to be a back barrier.[8, 49, 50] Simulations and experiments indicated that back contact barriers < 200 meV did not cause JV rollover at low temperature,[50] while the presently observed $E_{a,AS}$ were ≤ 100 meV. Further, the low temperature differential resistance (dV/dJ) did not correlate well with $E_{a,AS}$ (Figure S6 and Table 1). This is evidence that AS was unrelated to low temperature JV rollover, and unrelated to the back barrier. The observed AS behavior is not attributed to bulk acceptor defects because: (1) the $E_{a,AS}$ were 31 - 100 meV, which is too shallow for an acceptor (bulk $E_F$-$E_v$ from CV were 137 - 192 meV, assuming constant $N_v$ of $1.5 \cdot 10^{19}$[55]), (2) potential fluctuations in chalcopyrites cause the valence bands to be distributed over a range of energies, which results in broad $E_{defect}$-$E_v$ peaks,[53] while the AS signatures were sharp, (3) $E_{a,AS}$ varied substantially among samples, while $E_{defect}$-$E_v$ should be relatively constant among samples, and (4) the dielectric freeze out, such as that associated with bulk acceptors in $Cu_2ZnSn(Se,S)_4$/CdS solar cells,[56] was not observed. Similar AS behavior has also been observed in c-Si/a-Si solar cells, and was attributed to inversion strength at the interface ($E_F$-$E_v$ for the n-type c-Si absorber).[57] Observed AS signatures are therefore attributed to $E_c$-$E_F$ at the absorber/CdS interface. The x ~ 0.07 and 0.30 devices had stronger surface inversion, while the x ~ 0.22 had weaker surface inversion, relative to the x ~ 0 device. More study will be needed to understand these scattered data.

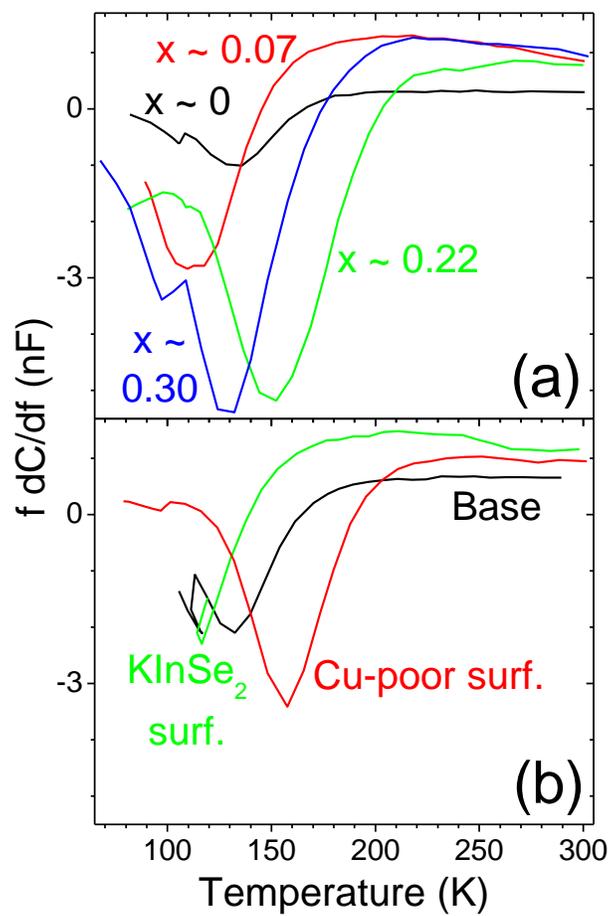

Figure S8. Admittance spectra for (a) bulk CKIS (grown at 500°C) with x ~ 0 (black), x ~ 0.07 (red), x ~ 0.22 (green), and x ~ 0.30 (blue) and (b) 600°C base (black), Cu-poor surface (red), and KInSe$_2$ surface (green) absorbers.

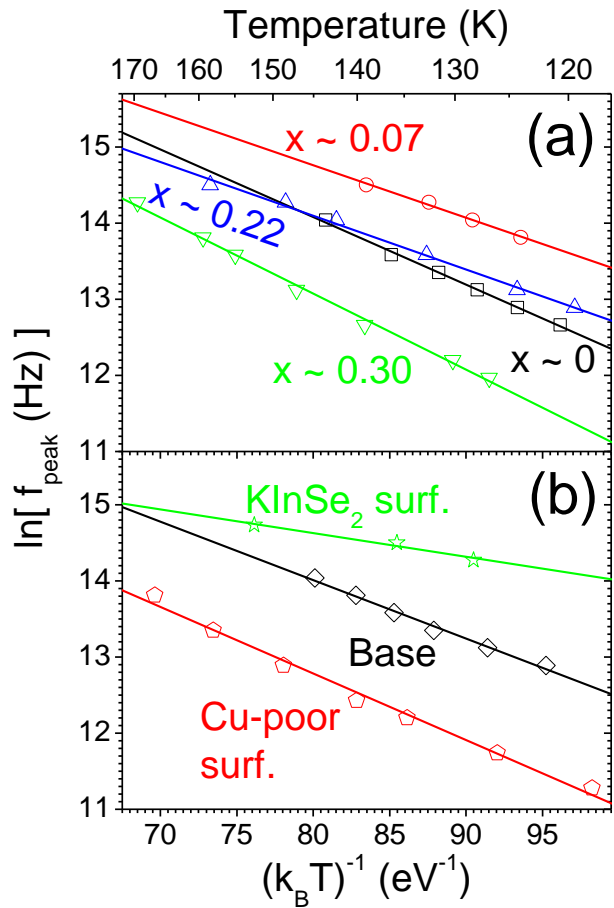

Figure 4. Arrhenius plot of the admittance peaks for (a) bulk CKIS (grown at 500°C) with x ~ 0 (black squares), x ~ 0.07 (red circles), x ~ 0.22 (green down triangles), and x ~ 0.30 (blue up triangles) and (b) 600°C base (black diamonds), Cu-poor surface (red pentagons), and KInSe$_2$ surface (green stars) absorbers. Lines are fits to the data.

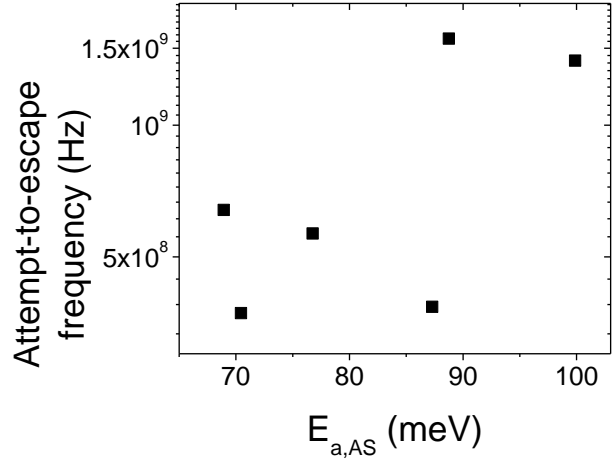

Figure S9. Attempt-to-escape frequency versus admittance spectroscopy activation energy for bulk CKIS (grown at 500°C) with x ~ 0, 0.07, 0.22, and x ~ 0.30, as well as 600°C base, Cu-poor surface, and KInSe$_2$ surface absorbers.

A technique was recently developed for quantifying recombination processes in a solar cell at $V_{OC}$.[52] This technique was previously applied to CIGS absorbers grown by three stage co-evaporation,[58] and to industry-fabricated cells with chalcogenized Cu(In,Ga)(Se,S)$_2$ absorbers and Zn(O,OH,S) buffers.[59] The technique requires built-in voltage, absorber majority carrier concentration, and depletion width from CV data. Absorber thickness, band gap, valence band density of states ($N_v$ of $1.5 \cdot 10^{19}$ cm$^{-3}$[55]), and intrinsic carrier concentration ($n_i$ of $5 \cdot 10^9$ cm$^{-3}$[55]) are also needed.[60] Published CKIS band gap values[44] were shifted to more closely match QE cutoffs (Figure S3), following previous CIGS recombination analyses.[58, 59] $V_{OC}(T)$ data can then be extrapolated to 0 K to yield $V_{OC}(T=0\ K)$ (Figure 3), and the ratio of absorber/buffer interface recombination to quasi-neutral, or bulk recombination. Finally, $V_{OC}(G)$ data are fit to a two-parameter equation to quantify room temperature interface, depletion region, and bulk recombination rates at $V_{OC}$ ($R_{interf.}$, $R_{depl.}$, and $R_{bulk}$; Figure S7). The results of this analysis technique are in Table 1. Additionally, ideal $V_{OC}$ ($V_{OC,id}$) was calculated for each $E_g$.[61] The deficit between actual and ideal $V_{OC}$ was then multiplied by relative recombination rate fraction (Table 1), e.g.:

$$D_{interf.} = \frac{(V_{OC} - V_{OC,id}) \cdot R_{interf.}}{R_{interf.} + R_{depl.} + R_{bulk}} \qquad (1)$$

This is a measure of deficit in photovoltage, or $V_{OC}$ losses resulting from interface, depletion region, and bulk recombination processes ($D_{interf.}$, $D_{depl.}$, $D_{bulk}$), respectively. While quantitative recombination rates at $V_{OC}$ are useful for understanding the operation of individual solar cells, they contain no information about relative recombination in solar cells with varied $V_{OC}$ and/or $E_g$. The presently proposed photovoltage deficit scales these recombination rates to actual performance losses, facilitating comparison among devices with unequal $V_{OC}$, absorbers with different $E_g$, and dissimilar PV technologies.

Recombination in the depletion region was negligible in all of the devices, in general agreement with previous analyses on CIGS devices.[58, 59] Increasing x from 0 to 0.07 reduced interface recombination, but at x ≥ 0.22, interface recombination dominated. This finding was consistent with $V_{OC}(T)$ data extrapolating to such low values at 0 K, relative to the band gaps of CKIS with x ≥ 0.22. The trend in interface recombination also resembled the trend in TRPL lifetimes. Increasing x composition decreased photovoltage deficits resulting from bulk recombination rates. However, apparent lifetimes did not reflect this trend. As TRPL excitation was near the absorbers' surfaces, the apparent lifetimes may have been affected by surface recombination. Two-photon excited TRPL will be needed to discern surface recombination velocities from bulk lifetimes in CKIS alloys.

## 2.2. High K/(K+Cu) at absorber surface

To probe the differences between bulk and surface effects of K, $CuInSe_2$ absorbers were grown at 600°C with: (1) homogenous cation evaporation rates ("baseline" or just "base"), (2) homogenous rates, except at the end of the growth, when the Cu rate was ramped down ("Cu-poor surface"), and (3) homogenous rates, except at the end, when the Cu rate was ramped down and the KF rate was ramped up ("$KInSe_2$ surface"). These films had very similar morphology (Figure S10). The $KInSe_2$ surfaces had in situ K/(K+Cu) of 0.30 to 0.92—conditions expected to produce $CuInSe_2$ and a large phase fraction of $KInSe_2$,[30, 45] albeit in a very thin surface layer (roughly 26 - 100 nm, depending on the particular growth). SIMS showed that the K was mostly confined to the absorber surface, but K levels throughout the film and at the Mo interface were also elevated (Figure S11). The PV performance trends with varied surface x composition, surface layer thickness, and growth temperature are in Figures S12 and S13. Increasing surface layer thickness (at constant x composition) and increasing x composition (at constant thickness) both had similar effects: efficiency, $V_{OC}$, and FF initially improved, but ultimately fell to worse than the baseline, while $J_{SC}$ went unchanged. With increasing thickness or x composition, the FF was reduced by series resistance, culminating in light JV rollover and crossover of the light and dark JV curves. This detrimental effect was sometimes reversed by a ~30 min light soak at 1 sun that improved $V_{OC}$, FF, and efficiency (e.g., Figure S14). These findings could relate to the high resistivity and photoconductivity of $KInSe_2$.[62] Growth temperature only affected $J_{SC}$, which was unrelated to the $KInSe_2$ surface (discussed below). These $KInSe_2$ surface absorbers had K- and In-enriched surfaces, similar to the reported effects of a KF PDT: more Cu-poor surfaces (equivalent to K- and In-enriched surfaces in the present work),[1, 7, 8, 10, 13-16, 18, 20, 24, 31-34] more K- and In-enriched surfaces,[10, 14, 18, 31, 32, 37-39] and wider surface band gaps (2.71 eV for $KInSe_2$[46, 62]).[14, 16, 24, 31, 32, 34] The absorbers with too much $KInSe_2$ at the surface (Figures S12, S13, and S14) also had blocking barriers similar to those observed in unsuccessful KF PDTs.[25] These similarities all serve as indirect evidence that $KInSe_2$ incidentally forms during KF PDTs.

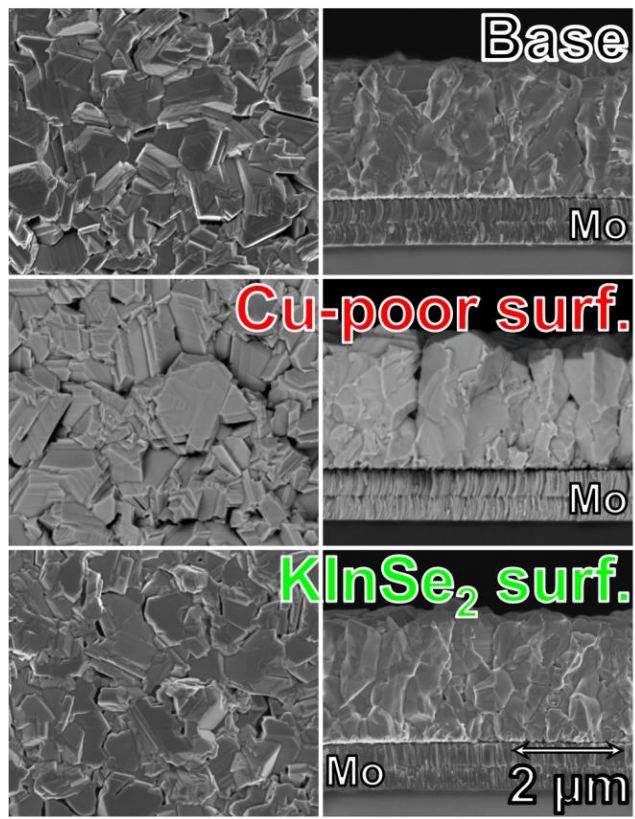

Figure S10. Plan view (left) and cross-sectional (right) SEM micrographs of baseline (top), Cu-poor surface (middle), and KInSe2 surface (bottom) absorbers grown at 600°C.

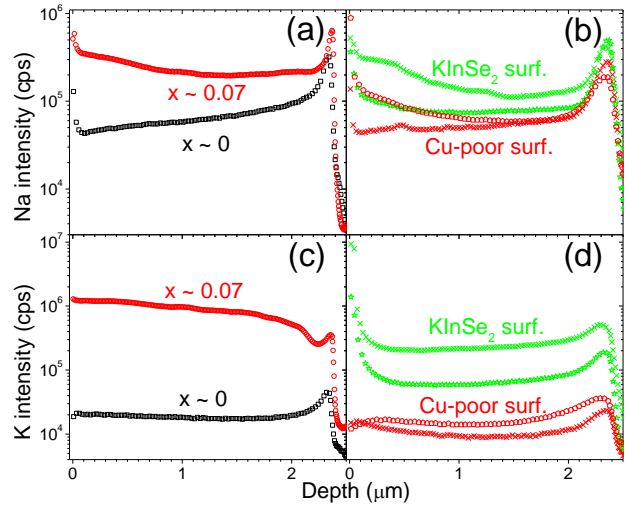

Figure S11. SIMS Na profiles for bulk CKIS (grown at 500°C) with x ~ 0 (black squares in (a)), x ~ 0.07 (red circles in (a)), 600°C-grown Cu-poor surfaces (red pentagons and crosses in (b)), and KInSe$_2$ surfaces (green stars and crosses in (b)), where K profiles are in (c) and (d) for the same scans. A depth of 0 is the absorbers' free surface; data is scaled so all films are equally thick.

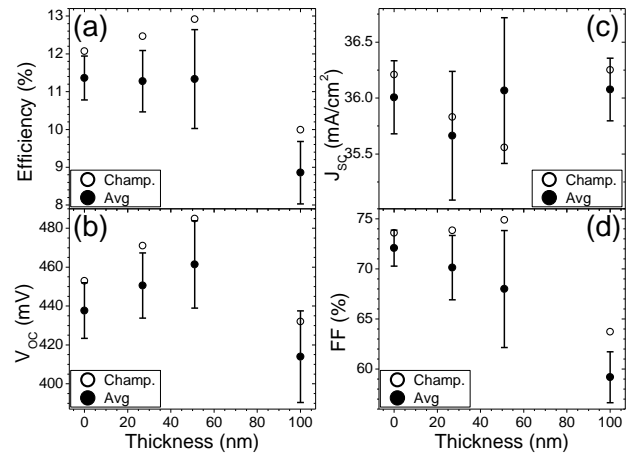

Figure S12. Champion (empty symbols) and average (filled symbols) efficiency (a), open-circuit voltage ($V_{OC}$) (b), short-circuit current density ($J_{SC}$) (c), and fill factor (FF) (d) versus KInSe$_2$ surface layer thickness, where x was 0.41 to 0.57, growth temperature was 500°C, and the intermediate layer was 67 nm. Standard deviations are shown as error bars.

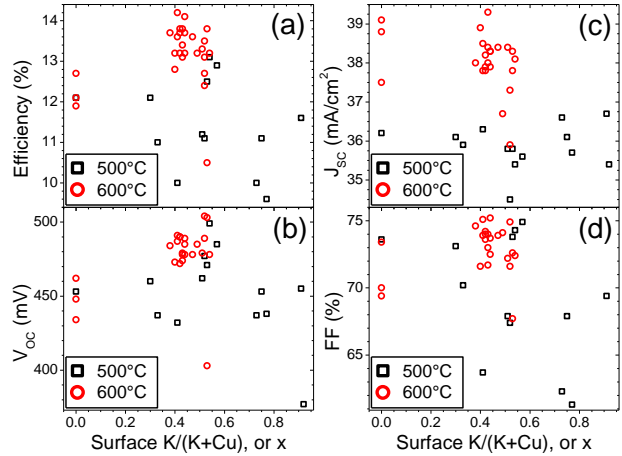

Figure S13. Champion 500°C (black squares) and 600°C (red circles) efficiency (a), open-circuit voltage ($V_{OC}$) (b), short-circuit current density ($J_{SC}$) (c), and fill factor (FF) (d) versus in situ $KInSe_2$ surface layer x composition, where other processing parameters were varied (e.g. intermediate and surface layers were 67 - 201 nm and 26 - 100 nm, respectively).

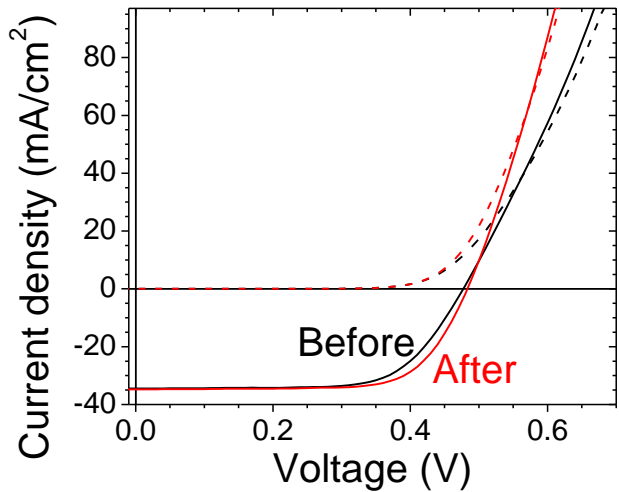

Figure S14. Dark (dashed) and light (solid) current density-voltage curves for a $KInSe_2$ surface device (grown at 500°C; x ~ 0.53; 201 nm intermediate layer; 38 nm surface layer) before (black) and after (red) a 68 min light soak (1 sun).

The champion devices for each absorber type were thoroughly characterized. Carrier concentration was slightly increased for the Cu-poor surface, and increased more for the $KInSe_2$ surface, in rough agreement with PL data (Table S1 and Figure S5). This may indicate that during KInSe$_2$ surface growth, K was incidentally incorporated into CuInSe$_2$ to increase carrier concentration—an unintended effect that may be difficult to eliminate in practice. The base TRPL lifetime was improved by 2x and 8x in the Cu-poor surface and KInSe$_2$ surface, respectively (Table S1 and Figure 2). Baseline absorbers grown at 600°C exhibited higher J$_{SC}$ and lower FF, relative to those grown at 500°C (Table 1). This may be related to a smoother absorber surface resulting from higher growth temperature (Figure S10), less CdS nucleation, and the resulting thinner CdS. Cu-poor surface absorbers had a small reduction in long wavelength QE (not shown), causing the smaller J$_{SC}$ (Table 1). The Cu-poor surface increased V$_{OC}$ and FF to enhance efficiency, while the KInSe$_2$ surface heightened these effects and introduced no J$_{SC}$ loss. After an anti-reflective coating was applied, the champion device with a KInSe$_2$ surface was officially measured at 14.9% efficient (Figure S15). It exhibited similar V$_{OC}$ and FF to the bulk x ~ 0.07 champion, although it had higher J$_{SC}$ (a direct result of its higher growth temperature). Low temperature JV rollover was diminished in the KInSe$_2$ surface sample (Figure S16), relative to the controls, and was further evidence of increased carrier concentration, likely due to CuInSe$_2$ incorporation of K. Similar to the bulk x ~ 0.22 and 0.30 champions, rollover in V$_{OC}$(T) at relatively high temperature was observed for the Cu-poor and KInSe$_2$ surface absorbers, relative to the 600°C baseline (Figure 3). The KInSe$_2$ surface V$_{OC}$(G) rolled over more strongly at low illumination, similar to the bulk x ~ 0.07 champion (Figure S7). The 600°C baseline and Cu-poor surface sample had similar surface inversion to the 500°C baseline (by AS; Figure 4 and Table 1). The KInSe$_2$ surface was substantially more inverted than all other samples, possibly a result of the CuInSe$_2$/KInSe$_2$/CdS bonds unique to that device. KF PDTs were also associated with increased inversion near the buffer interface,[1, 8, 10, 16, 33] further evidence that KInSe$_2$ forms during KF PDTs. AS on devices with different buffers and back contacts, as well as more extensive band alignment characterization will help to understand these trends. The baseline and Cu-poor surface samples grown at 600°C had more interface recombination than the 500°C baseline (Table 1). These samples also exhibited V$_{OC}$(T) rollover at higher temperatures—possible evidence of Fermi level pinning via interface recombination. The KInSe$_2$ surface had negligible interface recombination, and slightly increased bulk recombination, relative to the baseline and Cu-poor surface.

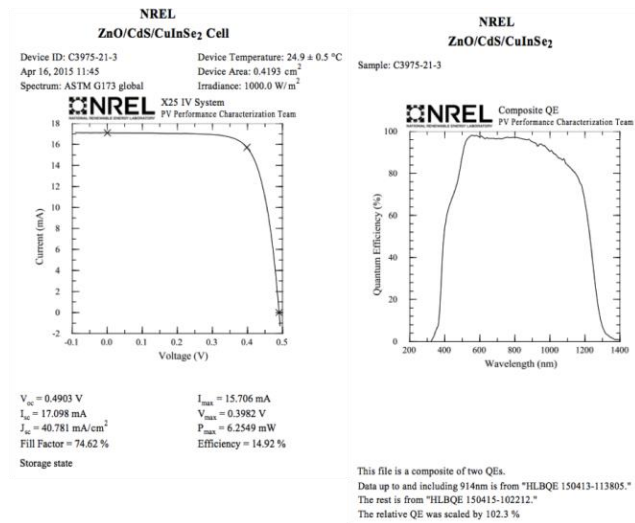

Figure S15. Officially measured current-voltage and external QE data for the KInSe$_2$ surface champion device.

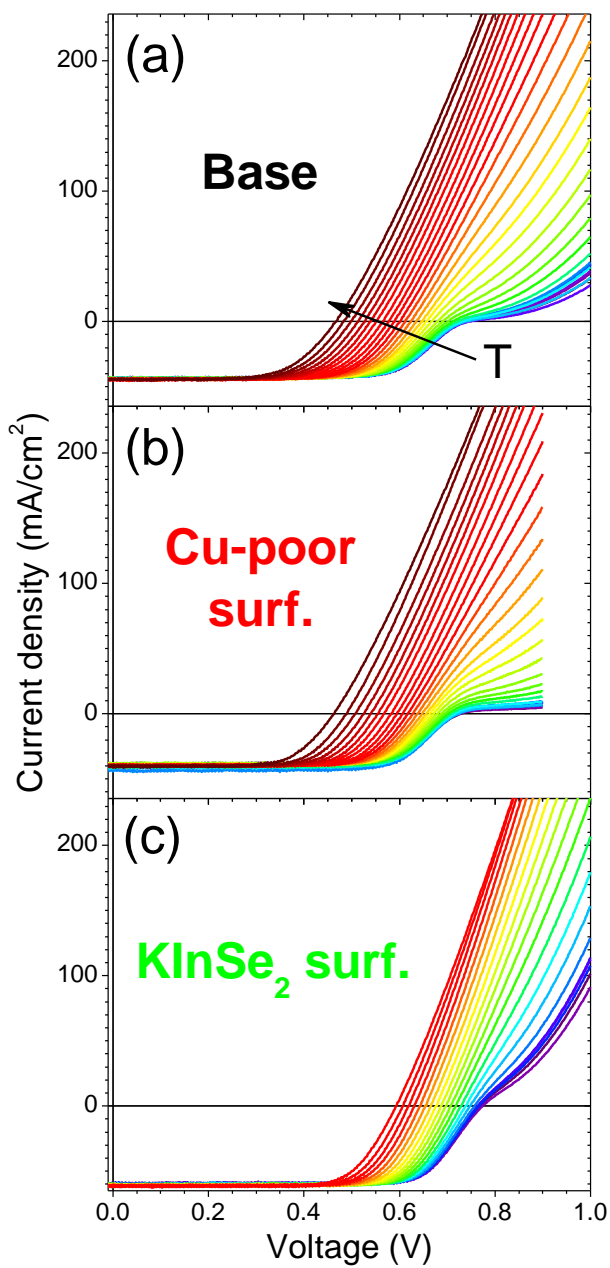

Figure S16. Temperature dependent JV at ~1 sun illumination for (a) baseline (106 - 290 K), (b) Cu-poor surface (79 - 302 K), and (c) KInSe$_2$ surface (114 - 298 K) absorbers grown at 600°C. Higher temperature curves are more red; lower temperature curves are more violet.

The bulk x ~ 0.07 and KInSe$_2$ surface champions thus achieved very similar PV performance through similar relative recombination rates, despite being grown with radically different processes. As mentioned above, both CKIS alloying and minor KInSe$_2$ could affect PV performance in the bulk CKIS alloy absorbers. The KInSe$_2$ surface champion could have likewise been affected by bulk CuInSe$_2$ film K-incorporation, but substantial CKIS alloying was unlikely at its high growth temperature.[30, 45] Future work to establish optimal bulk CKIS and KInSe$_2$ surface growth recipes on Na-poor (SiO$_2$/Mo) substrates[30] and at varied growth temperature[45] should help separate bulk absorber K-incorporation effects from KInSe$_2$ effects. The data suggest: (1) CKIS alloys with low K/(K+Cu) (~ 0.07) compositions at the absorber/buffer interface *decrease* recombination there, (2) CKIS alloys with high K/(K+Cu) (≥ 0.22) compositions at the absorber/buffer interface *increase* recombination there, and (3) mixed-phase CuInSe$_2$ + KInSe$_2$ with high K/(K+Cu) (> 0.22) compositions at the absorber/buffer interface *decrease* recombination there. High efficiencies are therefore only compatible high K/(K+Cu) compositions at the absorber/buffer interface when K is mostly incorporated into KInSe$_2$, *not* the chalcopyrite phase. The bulk CKIS x ~ 0.07 and KInSe$_2$ surface champions both achieved excellent performance through diminished interface recombination, a possible indication of similar beneficial mechanisms. However, the KInSe$_2$ surface had a much greater K composition at the absorber surface (Figure S11), exhibited superior apparent lifetimes (e.g. Table S1 and Figure 2), and had increased surface inversion (Table 1), relative to the bulk CKIS x ~ 0.07 champion. These properties indicate that the bulk CKIS alloy and KInSe$_2$ surface champions had improved performance through different mechanisms. Nanoscopic material characterization and density functional theory calculations will be needed to directly investigate this KInSe$_2$ interfacial layer, and the mechanism by which it can reduce recombination, but can also form a blocking barrier.

3. Conclusions

The effect of changing K/(K+Cu), or x composition in bulk CKIS alloy absorbers was studied. The bulk x ~ 0.07 film had improved efficiency, V$_{OC}$, and FF, relative to x ~ 0, 0.22, and 0.30. That film achieved an officially-measured 15.0% efficiency, matching the world record CuInSe$_2$ efficiency. Extensive electrical characterization revealed the mechanism of improvement to be reduced interface and bulk recombination, with no change in depletion region recombination. Higher x compositions had similar V$_{OC}$ to the x ~ 0.07 optimum, with diminished efficiency, FF, and J$_{SC}$—a result of increased interface recombination, relative to the x ~ 0 baseline. Next, the effect of confining K near the absorber/buffer interface was investigated: thin Cu-K-In-Se layers with higher K/(K+Cu), or x composition were evaporated onto CuInSe$_2$ films, conditions that produced a large phase fraction of KInSe$_2$.[30, 45] The thickness and x composition of this KInSe$_2$ surface layer were varied until PV performance was optimized at 75 nm and x ~ 0.41. The champion had enhanced efficiency (officially 14.9%), V$_{OC}$, and FF, relative to the baseline. This was a result of greatly reduced interface recombination, relative to the baseline and Cu-poor devices. Absorbers with KInSe$_2$ surfaces were similar to previous KF PDT reports,[1, 7, 8, 10, 13-16, 18, 20, 24, 31-34, 37-39] taken as evidence that KInSe$_2$ can form during a PDT. Both the bulk and surface growth processes greatly reduced interface recombination. However, the KInSe$_2$ surface had more K at the absorber surface, longer lifetimes, and greater inversion at the surface, relative to the bulk CKIS alloy champion. These data demonstrate that bulk K chalcopyrite-incorporation and absorber surface-confined KInSe$_2$, although difficult to experimentally separate, may benefit PV performance by different mechanisms.

5. Experimental Section

Most research on chalcopyrites utilizes the three stage co-evaporation deposition process to achieve high efficiencies at Ga/(Ga+In) compositions of 0.20 - 0.30.[1] Here, Ga was foregone so that observations could not be attributed to Ga/(Ga+In) changes. The three stage process was also avoided so that observations could not be attributed to altered cation profiles (potassium can affect cation diffusion[6, 40]). The deposition of ~2.5 μm CKIS absorbers with x of 0 - 0.30 was performed at 500°C on SLG/Mo substrates, as previously described.[44] KF evaporation rates below 0.7 Å/s were used to enhance control and reproducibility of the deposition process, as previously reported.[44] For KInSe$_2$ surface samples, the KF rate was ramped up and the Cu rate was ramped down over a period of 1 - 3 min near the end of the growth. This was followed by constant rate evaporation for 0.4 - 3.7 min, establishing surfaces with x of 0.30 - 0.92. Profilometry on the final film was used to infer individual layer thicknesses from in situ molar flux data, assuming constant density. The thicknesses of the compositionally-graded intermediate layers were varied from 47 to 201 nm, while the surface layers were 26 - 100 nm. Among these KInSe$_2$ surface samples, the champion had a 132 nm intermediate layer and a 75 nm surface layer, where the surface had K/(K+Cu), or x ~ 0.41. A control sample with a Cu-poor surface was prepared using that same cation profile, only with no KF added. For comparison, the bulk x ~ 0.07, bulk x ~ 0.22, bulk x ~ 0.30, and KInSe$_2$ surface films had 1.5, 5.1, 7.4, and 0.4 at.-% K by in situ measurement. Unlike typical KF PDT procedures,[1, 5-10, 14-16, 19, 24, 26, 31, 32, 34] absorbers were *not* rinsed before CBD. TRPL on bare absorbers was performed under low-injection conditions using a previously described system with 1.37 eV excitation and 0.92 - 1.31 eV detection.[63] High temperature vacuum anneals with Se over-pressure, XRD, XRF, SEM, and SIMS were performed on bare absorbers using previously described conditions.[44, 45, 64] Bulk absorber x compositions were estimated from XRD lattice parameters,[44] as CKIS alloy formation was shown to depend strongly on substrate surface Na[30] and temperature,[45] reducing the certainty of in situ x composition measurements. Solar cells were fabricated with 50 nm CdS, 100 nm i-ZnO, 120 nm Al:ZnO, 50 nm Ni, and 3 μm Al, as previously detailed.[65] JV and QE were performed under formerly reported conditions.[64] Room temperature CV measurements were performed at 10 kHz. Room temperature AS revealed a single signature for every device, and activation energies were extracted from least squares fit to the data. JV(T,G) measurements were performed at 68 - 302 K and $10^{-6}$ - $10^{0}$ suns using a neutral density filter. Electron-beam evaporation of 100 - 150 nm $MgF_2$ was applied to the best samples, and the two best devices were officially measured by the National Renewable Energy Laboratory PV Performance Characterization Team.

Acknowledgements

The authors would like to thank Carolyn Beall, Stephen Glynn, and Vincenzo LaSalvia for assistance with experiments, Clay DeHart for contact deposition, Karen Bowers for photolithography, Paul Ciszek for official performance measurement, Matt Young for SIMS, and Bobby To for SEM. The work was supported by the U.S. Department of Energy under FPACE contract DE-AC36-08GO28308.